\documentclass[12pt,thmsa]{article}
\usepackage{sw20aip}

\usepackage[T1]{fontenc}
\input{tcilatex}
\begin{document}

\author{Ezra T. Newman$^{1}$ and Gilberto Silva-Ortigoza$^{2}$ \\
$^{1}$Dept of Physcis and Astronomy, \\
Univ. of Pittsburgh, Pittsburgh, \\
PA 15260, USA \\
$^{2}$Facultad de Ciencias F\'{\i }sico Matem\'{a}ticas \\
de la Universidad Aut\'{o}noma de Puebla, \\
Apartado Postal 1152, Puebla, Pue., \\
M\'{e}xico}
\title{Tensorial Spin-s Harmonics }
\date{8.08.05 }
\maketitle

\begin{abstract}
We show how to define and go from the spin-s spherical harmonics to the
tensorial spin-s harmonics. These quantities, which are functions on the
sphere taking values as Euclidean tensors, turn out to be extremely useful
for many calculations in General Relativity. In the calculations, products
of these functions, with their needed decompositions which are given here,
often arise naturally.
\end{abstract}

\section{Introduction}

The use of the ordinary spherical harmonics, Y$_{lm}(\theta ,\varphi )$ (and
their associated vector and tensor harmonics) with all their properties,
their eigenvalue/eigenvector behavior, their orthonormality properties,
their use in solving various problems and equations in mathematical physics,
their interpretations in terms of multipole expansion have been ubiquitous
in theoretical physics. Several years ago a generalization of the ordinary
harmonics was developed\cite{NT,NP1} and referred to as the spin-s spherical
harmonics and denoted by

\[
_{(s)}Y_{lm}(\theta ,\varphi ) 
\]
with $s=0$ the ordinary spherical harmonics. They have proved to be very
useful in problems involving spin-s fields and have become an almost
essential tool in problems involving gravitational physics. Though the $
_{(s)}Y_{lm}$ are closely related to the Wigner D-matrices, $%
D_{l}^{mm^{\prime }}$ \{and to generalized versions of vector and tensor
spherical harmonics\} and can be derived from them, nevertheless it is the
particular form and specific properties of the $_{(s)}Y_{lm}$ that have been
of great use.

Recently, it turns out, that there are quantities that have been appearing
in many calculations that are closely related to the $_{(s)}Y_{lm}$, which
we refer to as the \textit{tensor spin-s harmonics }and are denoted by $%
Y_{(l)i.......k}^{(s)}.$ This note is devoted to a discussion of these new
harmonics.

There is a one-to-one correspondence between $Y_{(l)i.......k}^{(s)}$ and
the $_{(s)}Y_{lm}.$ The indices $i.....k$ indicate symmetric and trace-free
3-dimensional Euclidean tensors; the number of tensor indices equals $l.$
The number of independent components, in both $Y_{(l)i.......k}^{(s)}$ and $
_{(s)}Y_{lm},$ is the same; $N=2l+1$ and, in fact, the quantities $%
Y_{(l)i.......k}^{(s)}$ are just linear combinations of the $_{(s)}Y_{lm}$
and could in principle be written as 
\[
Y_{(l)i.......k}^{(s)}=\Sigma _{m}K_{(l)i.....k}^{(s)(m)}\,\cdot
\,_{{}(s)}Y_{lm}. 
\]
It is however difficult and a bit unwieldy to find the $%
K_{(l)a.....b}^{(s)(m)}$ and a more direct approach, namely to define them
independently, is considerably easier.

The value in using the $Y_{(l)i.......k}^{(s)}$ instead of the spin-s
spherical harmonics is two-fold. (i) They simply appear as they are defined
in many of the calculations and (ii) in non-linear theories very often
within calculations one finds that products of these tensor harmonics
automatically are there. They, in turn, must be decomposed via a
Clebsch-Gordon expansion. These expansions are most easily done with the $%
Y_{(l)i.......k}^{(s)}.$

In section II we will describe how the $Y_{(l)i.......k}^{(s)}$ arise in GR
and precisely how they are defined. In addition for clarity we give many
examples. In section III, we discuss how products [Clebsch-Gordon
expansions] are found and give an example from GR in section IV. The main
bulk of this note appears in appendices. Appendix A contains some useful
miscellaneous relations while Appendix B contains a table of Clebsch-Gordon
expansions of some of the most useful products.

\section{The Tensor Spin-s Harmonics}

\subsection{Notation \& Basics}

We begin with either Minkowski space or at an arbitrary point in a
Lorentzian space-time and take the metric as 
\begin{equation}
\eta _{ab}=diag[1,-1,-1,-1].  \label{eta}
\end{equation}

The standard null tetrad, parametrized by the complex stereographic angles ($
\zeta ,\overline{\zeta }$), with $\zeta =e^{i\phi }\cot \frac{\theta }{2},$
can be given by

\begin{eqnarray}
l^{a} &=&\frac{1}{\sqrt{2}(1+\zeta \overline{\zeta })}\left( 1+\zeta 
\overline{\zeta },\zeta +\overline{\zeta },-i(\zeta -\overline{\zeta }
),-1+\zeta \overline{\zeta }\right) ,  \label{nulltetrad1} \\
m^{a} &=&\frac{1}{\sqrt{2}(1+\zeta \overline{\zeta })}\left( 0,1-\overline{%
\zeta }^{2},-i(1+\overline{\zeta }^{2}),2\overline{\zeta }\right) ,
\label{2} \\
\overline{m}^{a} &=&\frac{1}{\sqrt{2}(1+\zeta \overline{\zeta })}\left(
0,1-\zeta ^{2},i(1+\zeta ^{2}),2\zeta \right) ,  \label{3} \\
n^{a} &=&\frac{1}{\sqrt{2}(1+\zeta \overline{\zeta })}\left( 1+\zeta 
\overline{\zeta },-(\zeta +\overline{\zeta }),i(\zeta -\overline{\zeta }
),1-\zeta \overline{\zeta }\right) ,
\end{eqnarray}
As ($\zeta ,\overline{\zeta })$ sweeps out the sphere, the null vector $
l^{a} $ sweeps out the light-cone and `drags along with it' the remainder of
the tetrad. From the tetrad we obtain the time-like and space-like vectors,

\begin{eqnarray}
t^{a} &\equiv &l^{a}+n^{a}=\sqrt{2}\left( 1,0,0,0\right)  \label{t} \\
c^{a} &\equiv &l^{a}-n^{a}=\frac{\sqrt{2}}{(1+\zeta \overline{\zeta })}
\left( 0,\zeta +\overline{\zeta },-i(\zeta -\overline{\zeta }),-1+\zeta 
\overline{\zeta }\right)  \label{c}
\end{eqnarray}
with norms 2 and -2 respectively.

The main tool or ingredient in our work will be the 3-dimensional
\{Euclidean\}vectors, $[i,j.....=1,2,3],$ obtained by projections that are
normal to $t^{a},$ i.e.,

\begin{eqnarray}
l_{i} &=&\frac{-1}{\sqrt{2}(1+\zeta \overline{\zeta })}\left( \zeta +%
\overline{\zeta },-i(\zeta -\overline{\zeta }),-1+\zeta \overline{\zeta }
\right) ,  \label{l} \\
m_{i} &=&\frac{-1}{\sqrt{2}(1+\zeta \overline{\zeta })}\left( 1-\overline{
\zeta }^{2},-i(1+\overline{\zeta }^{2}),2\overline{\zeta }\right) ,
\label{m} \\
\overline{m}_{i} &=&\frac{-1}{\sqrt{2}(1+\zeta \overline{\zeta })}\left(
1-\zeta ^{2},i(1+\zeta ^{2}),2\zeta \right) ,  \label{mbar} \\
n_{i} &=&-l_{i}=\frac{1}{\sqrt{2}(1+\zeta \overline{\zeta })}\left( (\zeta +%
\overline{\zeta }),-i(\zeta -\overline{\zeta }),-1+\zeta \overline{\zeta }%
\right) ,  \label{n} \\
c_{i} &=&l_{i}-n_{i}=\frac{-\sqrt{2}}{1+\zeta \overline{\zeta }}\left( \zeta
+\overline{\zeta },-i(\zeta -\overline{\zeta }),-1+\zeta \overline{\zeta }
\right) .  \label{c2}
\end{eqnarray}

In terms of $\theta $ and $\phi ,$

\begin{equation}
c_{i}=-\sqrt{2}(\cos \phi \sin \theta ,\sin \phi \sin \theta ,\cos \theta ).
\end{equation}
and hence $c_{i}$ is just $-\sqrt{2}$ times the unit Euclidean radial
vector. Note that we have used the Minkowski metric, (\ref{eta}) to raise
and lower even the Euclidean indices in Eqs.(\ref{1}) - (\ref{c2}) which
gives rise to the minus sign.

Some of the important algebraic properties of these vectors obtained by
direct calculation from their definitions, are 
\begin{eqnarray}
\delta _{ij} &=&\frac{1}{2}c_{i}c_{j}+\overline{m}_{i}m_{j}+m_{i}\overline{m}
_{j},  \label{delta} \\
c_{k} &=&-\sqrt{2}i\epsilon _{kij}m_{i}\overline{m}_{j},  \label{c3} \\
m_{k} &=&\frac{i}{\sqrt{2}}\epsilon _{kij}m_{i}c_{j},  \label{m3} \\
m_{j}\overline{m}_{k}-\overline{m}_{j}m_{k} &=&\frac{i}{\sqrt{2}}\epsilon
_{jki}c_{i},  \label{mmbar} \\
m_{k}c_{j}-m_{j}c_{k} &=&-i\sqrt{2}\epsilon _{kji}m_{i}.  \label{mc}
\end{eqnarray}

Noting that $c_{k},m_{k}$ and $\overline{m}_{k}$ have respectively
spin-weights (0,1, -1) and an $l$-value, $l=1,$ we have, either from the
definitions of $the$ differential operator edth acting on a spin-wt. s
function $\eta _{(s)},$ i.e., 
\begin{eqnarray}
\text{\dh }\eta _{(s)} &=&P_{0}^{1-s}\partial _{\zeta }(P_{0}^{s}\eta
_{(s)}),  \label{edth} \\
\overline{\text{\dh }}\eta _{(s)} &=&P_{0}^{1+s}\partial _{\overline{\zeta }%
}(P_{0}^{-s}\eta _{(s)}),  \label{edthb} \\
P_{0} &=&1+\zeta \overline{\zeta },  \label{P0}
\end{eqnarray}
or from the general eigenvalue relations \cite{NP1}

\begin{eqnarray}
\overline{\text{\dh }}\text{\dh }_{(s)}Y_{lm} &=&-(l-s)(l+s+1)_{(s)}Y_{lm},
\label{eign1} \\
\text{\dh }\overline{\text{\dh }}_{(s)}Y_{lm} &=&-(l+s)(l-s+1)_{(s)}Y_{lm},
\label{eign2}
\end{eqnarray}
the differential relations

\begin{eqnarray}
\text{\dh }c_{i} &=&2m_{i},  \label{eth1} \\
\overline{\text{\dh }}c_{i} &=&2\overline{m}_{i},  \label{eth2} \\
\text{\dh }\overline{m}_{i} &=&-c_{i},  \label{eth3} \\
\overline{\text{\dh }}m_{i} &=&-c_{i},  \label{eth4} \\
\overline{\text{\dh }}\text{\dh }c_{i} &=&-2c_{i},  \label{eth5} \\
\overline{\text{\dh }}\text{\dh }\overline{m}_{i} &=&-2\overline{m}_{i},
\label{eth6} \\
\text{\dh }\overline{\text{\dh }}m_{i} &=&-2m_{i}.  \label{eth7}
\end{eqnarray}

\subsection{Definition of $Y_{(l)i.......k}^{(s)}$}

The essential idea to define the tensor harmonics $Y_{(l)i.......k}^{(s)}$
is simply to use, in an appropriate way, tensor products of the three basic
Euclidean vectors, ($c_{i\text{ }},m_{i},\overline{m}_{i}).$ Since each of
them have the value $l=1$, in any product of $n$ terms the $l$-value of the
product is $l=n.$ The spin wt. of the product is given by the algebraic sum
of the spin wts. of the constituent vectors. In addition we require the
products to be symmetric trace-free Euclidean tensors.

The easiest way to do this is to first define the spin-wt. $s=l$ tensor
harmonic $Y_{(l)i.......k}^{(l)}$, \{with $l$ indices, i....k or $l$ factors 
$m_{k}$\} and $Y_{(s)i.......k}^{(-s)}$ by 
\begin{eqnarray}
Y_{(l)i.......k}^{(l)} &=&m_{i}m_{j}.....m_{k}  \label{Yss} \\
Y_{(l)i.......k}^{(-l)} &=&\overline{m}_{i}\overline{m}_{j}....\overline{m}%
_{k}.  \label{Ys-s}
\end{eqnarray}

First note that both are obviously symmetric and trace-free [since $%
m_{i}m_{j}\delta ^{ij}=0$] and that any derivatives will also be symmetric
and trace-free. Furthermore we recall that the edth operators
\[
\text{\dh }\ \text{and }\overline{\text{\dh }}
\]
are stepping operators, i.e., they add or subtract one to the value of $s.$

For positive values of $s,$ [$s=0,1,....,l],$ the $Y_{(l)i.......k}^{(s)}$
are defined by applying the operator 
\[
\overline{\text{\dh }}, 
\]
$l-s$ times to $Y_{(s)i.......k}^{(s)},$ i.e., 
\begin{equation}
Y_{(l)i.......k}^{(s)}=\overline{\text{\dh }}^{l-s}\{Y_{(l)i.......k}^{(l)}\}
\label{Ysl}
\end{equation}
and for negative values of $s,$ i.e., for [$0,-1,....,-l$], 
\begin{equation}
Y_{(l)i.......k}^{(-|s|)}=\text{\dh }^{l-|s|}\{Y_{(l)i.......k}^{(-l)}\}.
\label{Y-sl}
\end{equation}

Note that 
\[
Y_{(l)i.......k}^{(-|s|)}=\overline{Y}_{(l)i.......k}^{(s)}. 
\]
It is easy to show that the tensor harmonics satisfy the eigenvalue equations

\begin{eqnarray}
\overline{\text{\dh }}\text{\dh }Y_{(l)i.......k}^{(s)}
&=&-(l-s)(l+s+1)Y_{(l)i.......k}^{(s)},  \label{eign1*} \\
\text{\dh }\overline{\text{\dh }}Y_{(l)i.......k}^{(s)}
&=&-(l+s)(l-s+1)Y_{(l)i.......k}^{(s)}.  \label{eign2*}
\end{eqnarray}

\subsection{Examples}

In order to clarify these definitions and for later use, using Eqs.(\ref
{eth1})--(\ref{eth4}) and (\ref{delta}), we give several examples:

\textbf{l = 0:}\newline

\begin{equation}
Y_{0}^{0}=1.
\end{equation}
\qquad \qquad

\textbf{l = 1:}\newline

\begin{eqnarray}
Y_{1i}^{1} &=&m_{i},  \label{11} \\
Y_{1i}^{0} &=&\overline{\text{\dh }}Y_{1i}^{1}=\text{\dh }Y_{1i}^{-1}=-c_{i}
\label{10} \\
Y_{1i}^{-1} &=&\overline{m}_{i}.  \label{1-1}
\end{eqnarray}

\textbf{l = 2:}\newline

\begin{eqnarray}
Y_{2ij}^{2} &=&m_{i}m_{j},  \label{22} \\
Y_{2ij}^{1} &=&\overline{\text{\dh }}Y_{2ij}^{2}=-(c_{i}m_{j}+m_{i}c_{j}),
\label{21} \\
Y_{2ij}^{0} &=&\overline{\text{\dh }}Y_{2ij}^{1}=3c_{i}c_{j}-2\delta _{ij},
\label{20} \\
Y_{2ij}^{-1} &=&-(c_{i}\overline{m}_{j}+\overline{m}_{i}c_{j}),  \label{2-1}
\\
Y_{2ij}^{-2} &=&\overline{m}_{i}\overline{m}_{j}.  \label{2-2}
\end{eqnarray}

\textbf{l = 3:}\newline

\begin{eqnarray}
Y_{3ijk}^{3} &=&m_{i}m_{j}m_{k}  \label{l=3} \\
Y_{3ijk}^{2} &=&-(c_{i}m_{j}m_{k}+m_{i}c_{j}m_{k}+m_{i}m_{j}c_{k})  \nonumber
\\
Y_{3ijk}^{1} &=&-[\delta _{ij}m_{k}+\delta _{kj}m_{i}+\delta _{ik}m_{j}]+%
\frac{5}{2}[c_{i}c_{j}m_{k}+c_{k}c_{j}m_{i}+c_{i}c_{k}m_{j}]  \nonumber \\
Y_{3ijk}^{0} &=&6(\delta _{ij}c_{k}+\delta _{ik}c_{j}+\delta
_{kj}c_{i})-15c_{i}c_{j}c_{k}  \nonumber
\end{eqnarray}

\textbf{l = 4:}\newline

\begin{eqnarray}
Y_{4ijkl}^{4} &=&m_{i}m_{j}m_{k}m_{l}  \label{l=4} \\
Y_{4ijkl}^{3}
&=&-c_{i}m_{j}m_{k}m_{l}-m_{i}c_{j}m_{k}m_{l}-m_{i}m_{j}c_{k}m_{l}-m_{i}m_{j}m_{k}c_{l}
\nonumber \\
Y_{4ijkl}^{2} &=&2[c_{i}c_{l}m_{j}m_{k}-\overline{m}_{l}m_{i}m_{j}m_{k}-%
\overline{m}_{k}m_{i}m_{j}m_{l}-\overline{m}_{j}m_{i}m_{k}m_{l}-\overline{m}
_{i}m_{j}m_{k}m_{l}  \nonumber \\
&&+c_{j}m_{k}(c_{l}m_{i}+c_{i}m_{l})+c_{k}(c_{l}m_{i}m_{j}+(c_{j}m_{i}+c_{i}m_{j})m_{l})]
\nonumber \\
Y_{4ijkl}^{1} &=&6[\overline{m}_{l}c_{k}m_{i}m_{j}+\overline{m}
_{k}c_{l}m_{i}m_{j}+\overline{m}_{j}c_{l}m_{i}m_{k}+\overline{m}
_{i}c_{l}m_{j}m_{k}+\overline{m}_{j}c_{k}m_{i}m_{l}  \nonumber \\
&&+\overline{m}_{i}c_{k}m_{j}m_{l}+c_{i}(-c_{k}c_{l}m_{j}+\overline{m}
_{l}m_{j}m_{k}+(\overline{m}_{k}m_{j}+\overline{m}_{j}m_{k})m_{l})  \nonumber
\\
&&+c_{j}(-c_{i}c_{l}m_{k}+\overline{m}_{l}m_{i}m_{k}+\overline{m}
_{k}m_{i}m_{l}+\overline{m}_{i}m_{k}m_{l}-c_{k}(c_{l}m_{i}+c_{i}m_{l}))] 
\nonumber \\
Y_{4ijkl}^{0} &=&105c_{i}c_{j}c_{k}c_{l}+12[\delta _{ik}\delta _{jl}+\delta
_{il}\delta _{jk}+\delta _{ij}\delta _{kl}]  \nonumber \\
&&-30[\delta _{ij}c_{k}c_{l}+\delta _{ik}c_{j}c_{l}+\delta
_{il}c_{k}c_{j}+\delta _{kj}c_{i}c_{l}+\delta _{lj}c_{k}c_{i}+\delta
_{kl}c_{i}c_{j}]  \nonumber
\end{eqnarray}

\section{Products of the $Y_{(l)i.......k}^{(s)}$}

Very often in doing detailed calculations in general relativity due to the
non-linearity one has to deal with products of different $%
Y_{(l)i.......k}^{(s)}.$ Usually they involve only small values of both $s$
and $l$, most often from 0 to 3 or 4. In principle, by using the products of
the Wigner $D$-functions, i.e., Clebsch-Gordon expansions, one could work
out the $Y_{(l)i.......k}^{(s)}$ products. In practice since the conversion
of the $Y_{(l)i.......k}^{(s)}$ to the $D_{mm^{\prime }}^{l}$ is quite
complicated - with various conventions - and we are most often interested in
the low values of $s$ and $l,$ it is easier to do the calculation of the
expansion directly.

First, instead of using ($c_{i\text{ }},m_{i},\overline{m}_{i}),$ we use 
\[
(Y_{1i}^{0},Y_{1i}^{1}Y_{1i}^{-1})=(-c_{i\text{ }},m_{i},\overline{m}_{i}). 
\]

The simplest and most important product relations are found most easily by
direct calculations using the definitions, (\ref{m}),(\ref{mbar})and (\ref
{c2}), of ($c_{i\text{ }},m_{i},\overline{m}_{i})$ yielding

\begin{eqnarray}
Y_{1i}^{1}Y_{1j}^{-1}-Y_{1j}^{1}Y_{1i}^{-1} &=&-\frac{i}{\sqrt{2}}\epsilon
_{ijk}Y_{1k}^{0},  \label{ij-ji} \\
Y_{1j}^{1}Y_{1k}^{0}-Y_{1k}^{1}Y_{1j}^{0} &=&i\sqrt{2}\epsilon
_{jki}Y_{1i}^{1}.  \label{ij-ji/2}
\end{eqnarray}
To find the products $Y_{1k}^{1}Y_{1j}^{0},$ $Y_{1k}^{1}Y_{1j}^{-1}$ and $%
Y_{1k}^{0}Y_{1j}^{0}$ one writes them out as a sum of tensor terms from $l=0$
to $l=2$ and applies the operator 
\[
\overline{\text{\dh }}\dh 
\]
\ several times using the eigenvalue equations (\ref{eign1*}) and (\ref
{eign2*}) to determine the coefficients.

For example $Y_{1j}^{0}Y_{1k}^{0}$ could be written out as

\begin{equation}
Y_{1j}^{0}Y_{1k}^{0}=AY_{0}^{0}+B^{j}Y_{1j}^{0}+C^{jk}Y_{2jk}^{0}  \label{A}
\end{equation}
with $A,B^{i}$ and $C^{ij}$ to be determined, and then from the
eigenfunction relations, we have 
\begin{eqnarray}
\overline{\text{\dh }}\text{\dh [}Y_{1j}^{0}Y_{1k}^{0}]
&=&-2B^{j}Y_{1j}^{0}-6C^{jk}Y_{2jk}^{0}  \label{B} \\
\overline{\text{\dh }}\text{\dh }\overline{\text{\dh }}\text{\dh [}%
Y_{1j}^{0}Y_{1k}^{0}] &=&4B^{j}Y_{1j}^{0}+36C^{jk}Y_{2jk}^{0}.  \label{C}
\end{eqnarray}
Since the left sides of (\ref{A}),(\ref{B}) and (\ref{C}) are known we can
evaluate the $A,B^{i}$ and $C^{ij}$ yielding 
\begin{equation}
Y_{1i}^{0}Y_{1j}^{0}=\frac{2}{3}\delta _{ij}+\frac{1}{3}Y_{2ij}^{0}.
\label{D}
\end{equation}

In a similar manner, with the help of (\ref{ij-ji}) and (\ref{ij-ji/2}), we
obtain 
\begin{eqnarray}
Y_{1i}^{1}Y_{1j}^{0} &=&\frac{i}{\sqrt{2}}\epsilon _{ijk}Y_{1k}^{1}+\frac{1}{
2}Y_{2ij}^{1}  \label{E} \\
Y_{1i}^{1}Y_{1j}^{-1} &=&\frac{1}{3}\delta _{ij}-\frac{i\sqrt{2}}{4}\epsilon
_{ijk}Y_{1k}^{0}-\frac{1}{12}Y_{2ij}^{0}.  \label{F}
\end{eqnarray}

In principle all products expansions can be found in this manner. As the
details can become quite tedious we simply list the most important ones in
the Appendix B.

\section{Application: The Robinson-Trautman Equation}

Though there are many applications of these results that will be given
elsewhere, here we will show how the tensor harmonics can be used to
approximate solutions to the Robinson-Trautman equation\cite{RT}, the final
equation that determines the type II, non-twisting metrics. Our purpose, in
this example, is simply to illustrate how the tensor harmonics and their
products enter into GR calculations and so we will not be concerned with the
certain details.

I. Robinson and A. Trautman in their investigation of algebraically special
vacuum metrics with non-twisting principle null vector found that the
algebraically special type II metrics could be reduced to the single partial
differential equation for a `mass parameter', $\chi (\tau )$, as a function
of the `time' parameter and a function 
\[
P(\tau ,\zeta ,\overline{\zeta })=V(\tau ,\zeta ,\overline{\zeta })(1+\zeta 
\overline{\zeta })\equiv V(\tau ,\zeta ,\overline{\zeta })P_{0}, 
\]
a time dependent conformal factor for a two-surface metric.

\begin{remark}
The function $\chi (\tau )$ though not the Bondi mass is a close relative.
The Bondi mass is given, up to a numerical factor, by $M_{B}=\chi (\tau
)W(\tau ),$ where 
\begin{equation}
W(\tau )=\int \frac{d\zeta d\overline{\zeta }}{V^{3}(1+\zeta \overline{\zeta 
})^{2}}  \label{W}
\end{equation}

One can easily show\cite{NPos} from Eq.(\ref{RT1}) that $M_{B}^{\prime }<0,$
i.e., the Bondi mass loss theorem.
\end{remark}

The Robinson/Trautman equation could be rewritten\cite{NPos}, using the edth
notation, (\ref{edth}), as 
\begin{mathletters}
\begin{equation}
\chi ^{^{\prime }}-3\frac{V^{^{\prime }}}{V}\chi =V^{3}[\text{\dh }\overline{%
\text{\dh }}\text{\dh }\overline{\text{\dh }}V+2\text{\dh }\overline{\text{
\dh }}V]-V^{2}(\text{\dh }^{2}V)(\overline{\text{\dh }}^{2}V)  \label{RT1}
\end{equation}
with ( $^{\prime }$ ) meaning the $\tau $-derivative. They showed that by
choosing a different time parameter $\tau \rightarrow \tau ^{*}=F(\tau )$
and rescaling the $V$ one could then make 
\end{mathletters}
\[
\chi ^{\prime }=0 
\]
thereby simplifying the equation. We however will use the reparametrizaion
freedom in a different way: we choose it to make the leading term in $V$ to
be one and thereby allow $\chi $ to be time dependent. In particular we
assume that $V$ takes the form of the tensor harmonic expansion with the $
l=0 $ term unity:

\begin{equation}
V=1-\frac{1}{2}\eta ^{i\,}Y_{1i}^{0}+\eta ^{ij}Y_{2ij}^{0}+.....  \label{V}
\end{equation}
with $\eta ^{ij}$ symmetric and trace-free.

\begin{remark}
: In other publications, and for specific reasons, we have chosen the $\eta $
's as $\tau $-derivatives of other functions, i.e., $\eta =\xi ^{\prime }$.
In the present work this is not necessary.
\end{remark}

Our object is to try to find approximate solutions to (\ref{RT1}) by
assuming that, in an expansion near the Schwarzschild solution, $\chi $ is
zero order while $\eta ^{i\,}$ and $\eta ^{i\,j}$ are respectively first and
second order. Note that the order of the $\tau $-derivatives is at this
stage not know but be determined by the differential equation, (\ref{RT1}).
Furthermore we will truncate the harmonic series at two, the quadrupole
term. The expression for $V,$ (\ref{V}), will be substituted into (\ref{RT1}
) and expanded up to the $l=2.$ This yields three different evolution
equations, $[l=0,1,2]$, for the $\tau $-derivatives of $\chi $, $\eta
^{i\,}\ $and $\eta ^{ij}.$ For a reason made clear later we will keep terms
up to fourth order, even though most will eventually be discarded.

As a preliminary to the substitution of $V$ into (\ref{RT1}) we calculate
from the eigenvalue equations, \textbf{(}\ref{eign1*}\textbf{) }and (\ref
{eign2*}),\textbf{\ }the following relations:

\begin{equation}
\text{\dh }\overline{\text{\dh }}V=-\frac{1}{2}\eta ^{i\,}\text{\dh }%
\overline{\text{\dh }}Y_{1i}^{0}+\eta ^{ij}\text{\dh }\overline{\text{\dh }}%
Y_{2ij}^{0}=\eta ^{i\,}Y_{1i}^{0}-6\eta ^{ij}Y_{2ij}^{0},  \label{aux1}
\end{equation}

\begin{equation}
\text{\dh }\overline{\text{\dh }}\text{\dh }\overline{\text{\dh }}V=-2\eta
^{i\,}Y_{1i}^{0}+36\eta ^{ij}Y_{2ij}^{0},  \label{aux2}
\end{equation}

\begin{equation}
\text{\dh }\overline{\text{\dh }}\text{\dh }\overline{\text{\dh }}V+2\text{
\dh }\overline{\text{\dh }}V=24\eta ^{ij}Y_{2ij}^{0},  \label{aux3}
\end{equation}

\begin{equation}
\text{\dh }^{2}V=\text{\dh }^{2}[1-\frac{1}{2}\eta ^{i\,}Y_{1i}^{0}+\eta
^{ij}Y_{2ij}^{0}]=\eta ^{ij}\text{\dh }^{2}Y_{2ij}^{0}=24\eta
^{ij}Y_{2ij}^{2},  \label{aux4}
\end{equation}

\begin{equation}
\overline{\text{\dh }}^{2}V=24\eta ^{ij}Y_{2ij}^{-2},  \label{aux5}
\end{equation}

\begin{equation}
(\text{\dh }^{2}V)(\overline{\text{\dh }}^{2}V)=(24)^{2}\eta ^{ij}\eta
^{kl}Y_{2ij}^{2}Y_{2kl}^{-2},  \label{aux6}
\end{equation}

Using these relations, Eq.(\ref{RT1}) can be written as 
\begin{mathletters}
\begin{equation}
V\chi ^{^{\prime }}-3V^{^{\prime }}\chi -V^{4}[24\eta
^{ij}Y_{2ij}^{0}+...]+V^{3}[(24)^{2}\eta ^{ij}\eta
^{kl}Y_{2ij}^{2}Y_{2kl}^{-2}+...]=0  \label{RT2}
\end{equation}
By expanding $V^{4}$ as

\end{mathletters}
\begin{eqnarray}
V^{4} &=&[1-\frac{1}{2}\eta ^{i\,}Y_{1i}^{0}+\eta ^{ij}Y_{2ij}^{0}]^{4}
\label{V^4} \\
&=&1-2\eta ^{i\,}Y_{1i}^{0}+\frac{3}{2}\eta ^{i\,}\eta
^{j\,}Y_{1i}^{0}Y_{1j}^{0}+4\eta ^{ij}Y_{2ij}^{0}+...  \nonumber
\end{eqnarray}
and using the product

\begin{equation}
Y_{1i}^{0}Y_{1j}^{0}=\frac{2}{3}\delta _{ij}+\frac{1}{3}Y_{2ij}^{0}
\label{P1}
\end{equation}
we have 
\begin{equation}
V^{4}=1-2\eta ^{i\,}Y_{1i}^{0}+\eta ^{i\,}\eta ^{i\,}+(\frac{1}{2}\eta
^{i\,}\eta ^{j\,}+4\eta ^{ij})Y_{2ij}^{0}+...  \label{V^.2}
\end{equation}

Up to fourth-order, Eq.(\ref{RT2}) reduces to

\begin{eqnarray}
&&-\chi ^{^{\prime }}+\frac{1}{2}(\chi ^{^{\prime }}\eta ^{i\,}-3\chi \eta
^{i\,\prime })Y_{1i}^{0}+[-\chi ^{^{\prime }}\eta ^{ij}+3\chi ^{^{\prime
}}\eta ^{ij\prime }+24(1+\eta ^{k}\eta ^{k})\eta ^{ij}]Y_{2ij}^{0}
\label{RT3} \\
&&-48\eta ^{i\,}\eta ^{kl}Y_{1i}^{0}Y_{2kl}^{0}+(12\eta ^{i\,}\eta
^{j\,}+96\eta ^{ij})\eta ^{kl}Y_{2ij}^{0}Y_{2kl}^{0}  \nonumber \\
&&-(24)^{2}\eta ^{ij}\eta ^{kl}Y_{2ij}^{2}Y_{2kl}^{-2}=0  \nonumber
\end{eqnarray}

From the product relations

\[
Y_{1i}^{0}Y_{2jk}^{0}=-\frac{4}{5}\delta _{kj}Y_{1i}^{0}+\frac{6}{5}(\delta
_{ij}Y_{1k}^{0}+\delta _{ik}Y_{1j}^{0})+\frac{1}{5}Y_{3ijk}^{0} 
\]
and those for $Y_{2ij}^{0}Y_{2kl}^{0}$ and $Y_{2ij}^{2}Y_{2kl}^{-2},$ given
in the appendix B, (and using the symmetry and trace-free properties of $%
\eta ^{ij}$), we have after a lengthy calculation that Eq.(\ref{RT3})
becomes, to fourth-order and to the $l=2$ harmonic,

\begin{eqnarray}
&&-\chi ^{\prime }+\frac{1}{2}\{\chi ^{\prime }\eta ^{i\,}-3\chi \eta
^{i\,\prime }\}Y_{1i}^{0}+\{-\chi ^{^{\prime }}\eta ^{ij}+3\chi \eta
^{ij\prime }+24(1+\eta ^{k\,}\eta ^{k})\eta ^{ij}\}Y_{2ij}^{0}  \label{RT4}
\\
&&+48\{\frac{6}{5}(\eta ^{i\,}\eta ^{j\,}+8\eta ^{ij})\eta ^{ij}-\frac{48}{7}
\{[\frac{1}{24}\eta ^{k\,}\eta ^{k\,}\eta ^{ij}-(\frac{1}{8}\eta ^{i\,}\eta
^{k\,}+\eta ^{ik})\eta ^{kj}]Y_{2ij}^{0}\}\} \\
&&-48\{\frac{12}{5}\eta ^{j\,}\eta ^{ji}Y_{1i}^{0}\}-(24)^{2}\{\frac{1}{5}
\eta ^{ij}\eta ^{ij}-\frac{1}{7}\eta ^{ik}\eta ^{jk}Y_{2ij}^{0}\}=0. 
\nonumber
\end{eqnarray}

From the $l=0,1,2$ harmonics this is equivalent to

\begin{equation}
-\chi ^{^{\prime }}+\frac{(24)^{2}}{5}(\frac{1}{2}\eta ^{i\,}\eta
^{j\,}+3\eta ^{ij})\eta ^{ij}=0,  \label{RT^1}
\end{equation}

\begin{equation}
3\chi \eta ^{i\,\prime }-\chi ^{^{\prime }}\eta ^{i\,}+\frac{2(24)^{2}}{5}
\eta ^{j\,}\eta ^{ji}=0,  \label{RT^2}
\end{equation}

\begin{eqnarray}
&&-\chi ^{\prime }\eta ^{ij}+3\chi \eta ^{ij\prime }+24\eta ^{ij}+\frac{12(6)%
}{7}\eta ^{k\,}\eta ^{k\,}\eta ^{ij}  \label{RT^3} \\
&&+SymTrFr(ij)\left( \frac{(24)^{2}}{7}(\frac{1}{2}\eta ^{i\,}\eta
^{k\,}+5\eta ^{ik})\eta ^{kj}\right) =0,  \nonumber
\end{eqnarray}
where $SymTrFr(ij)$ means symmetrize and remove the trace.

These three equations will be used in two different ways: (i) They will
first be used to determine the order of the $\tau $-derivatives and
consequently simplify the equations and (ii) then integrate them.

From the first we see that $\chi ^{^{\prime }}$ is fourth order 
\begin{equation}
\chi ^{^{\prime }}=\frac{(24)^{2}}{5}(\frac{1}{2}\eta ^{i\,}\eta
^{j\,}+3\eta ^{ij})\eta ^{ij}=0.  \label{RT^*}
\end{equation}
Using this in the second equation and keeping only the leading term, we have
that $\eta ^{i\,\prime }$ is third order, 
\begin{equation}
\chi \eta ^{i\,\prime }=-\frac{(16)(24)}{5}\eta ^{j\,}\eta ^{ji}.
\label{RT2^*}
\end{equation}
while the third equation for $\eta ^{ij\,\prime },$ is second order, 
\begin{equation}
\chi \eta ^{ij\prime }=-8\eta ^{ij}.  \label{RT3^*}
\end{equation}
These equations, by taking them in the reverse order and treating $\chi =$
constant in the 2$^{nd}$ and 3$^{rd}$ equations, can be integrated as

\begin{equation}
\eta ^{ij}=\eta _{0}^{ij}e^{-8\chi ^{-1}\tau }.  \label{n_ij}
\end{equation}
The second equation can be integrated exactly with the solutions depending
on an exponential of the exponential $e^{-8\chi ^{-1}\tau },$ i.e., it
involves functions of the form 
\[
e^{\exp [-8\chi ^{-1}\tau ]} 
\]
which goes to one rapidly so here we simply take $\eta ^{i\,}$ as constant, 
\[
\eta ^{i\,}=\eta _{0}^{i\,}. 
\]
[The same result could be obtained by neglecting the third order term.] The
first equation then simply integrates to the form

\begin{equation}
\chi =\chi _{0}+Ke^{-16\chi ^{-1}\tau }.  \label{Xi}
\end{equation}

Since here we are not interested in the details of the Robinson-Trautman
metric, but only to illustrate the use of the tensor harmonics, we will not
give the detailed expressions for these quantities. We simply mention that
using Eqs.(\ref{W}) and (\ref{Xi}), we can obtain the evolution of the Bondi
mass.

\section{Acknowledgments}

This material is based upon work (partially) supported by the National
Science Foundation under Grant No. PHY-0244513. Any opinions, findings, and
conclusions or recommendations expressed in this material are those of the
authors and do not necessarily reflect the views of the National Science.
E.T.N. thanks the NSF for this support. G.S.O. acknowledges the financial
support from CONACyT through Grand No.44515-F, VIEP-BUAP through Grant No.
II161-04/EXA and Sistema Nacional de Investigadores, (M\'{e}xico).

\section{Appendices}

\subsection{miscellaneous relations}

\textbf{l = 1:}\newline

\begin{eqnarray}
Y_{1i}^{0} &=&\overline{\text{\dh }}Y_{1i}^{1}=-c_{i} \\
Y_{1i}^{0} &=&\text{\dh }Y_{1i}^{-1}=-c_{i} \\
\text{\dh }Y_{1i}^{0} &=&-2m_{i}=-2Y_{1i}^{1} \\
\overline{\text{\dh }}Y_{1i}^{0} &=&-2\overline{m}_{i}=-2Y_{1i}^{-1} \\
\overline{\text{\dh }}\text{\dh }Y_{1i}^{0} &=&-2Y_{1i}^{0} \\
\text{\dh }Y_{1i}^{1} &=&0\,\,\Rightarrow \overline{\text{\dh }}\text{\dh }
Y_{1i}^{1}=0 \\
\overline{\text{\dh }}Y_{1i}^{-1} &=&0\,\,\Rightarrow \text{\dh }\overline{%
\text{\dh }}Y_{1i}^{-1}=0, \\
\text{\dh }\overline{\text{\dh }}Y_{1i}^{1} &=&-2Y_{1i}^{1}, \\
\overline{\text{\dh }}\text{\dh }Y_{1i}^{-1} &=&-2Y_{1i}^{-1}.
\end{eqnarray}

\textbf{l = 2:}\newline

\begin{eqnarray}
Y_{2ij}^{1} &=&\overline{\text{\dh }}Y_{2ij}^{2}=-(c_{i}m_{j}+m_{i}c_{j}) \\
\text{\dh }Y_{2ij}^{1} &=&\text{\dh }\overline{\text{\dh }}
Y_{2ij}^{2}=-4m_{i}m_{j}=-4Y_{2ij}^{2} \\
Y_{2ij}^{0} &=&\overline{\text{\dh }}^{2}Y_{2ij}^{2}=3c_{i}c_{j}-2\delta
_{ij} \\
\text{\dh }Y_{2ij}^{0} &=&6(m_{i}c_{j}+c_{i}m_{j})=-6Y_{2ij}^{1} \\
\text{\dh }\overline{\text{\dh }}Y_{2ij}^{0} &=&-6Y_{2ij}^{0}
\end{eqnarray}

\textbf{l = 3:}\newline

\begin{eqnarray}
Y_{3ijk}^{2} &=&\overline{\text{\dh }}%
Y_{3ijk}^{3}=-(c_{i}m_{j}m_{k}+m_{i}c_{j}m_{k}+m_{i}m_{j}c_{k}) \\
Y_{3ijk}^{1} &=&\overline{\text{\dh }}
Y_{3ijk}^{2}=2[c_{i}c_{j}m_{k}+c_{i}m_{j}c_{k}+m_{i}c_{j}c_{k}  \nonumber \\
&&-\overline{m}_{i}m_{j}m_{k}-m_{i}\overline{m}_{j}m_{k}-m_{i}m_{j}\overline{
m}_{k}] \\
Y_{3ijk}^{1} &=&4[\delta _{ij}m_{k}+\delta _{kj}m_{i}+\delta _{ik}m_{j}] 
\nonumber \\
&&-10[m_{i}\overline{m}_{j}m_{k}+\overline{m}_{i}m_{j}m_{k}+m_{i}m_{j}%
\overline{m}_{k}] \\
Y_{3ijk}^{1} &=&-[\delta _{ij}m_{k}+\delta _{kj}m_{i}+\delta _{ik}m_{j}]+%
\frac{5}{2}[c_{i}c_{j}m_{k}+c_{k}c_{j}m_{i}+c_{i}c_{k}m_{j}] \\
Y_{3ijk}^{0} &=&\overline{\text{\dh }}Y_{3ijk}^{1}=6(\delta
_{ij}c_{k}+\delta _{ik}c_{j}+\delta _{kj}c_{i})-15c_{i}c_{j}c_{k} \\
\text{\dh }Y_{3ijk}^{2} &=&\text{\dh }\overline{\text{\dh }}
Y_{3ijk}^{3}=-6m_{i}m_{j}m_{k}=-6Y_{3ijk}^{3} \\
\text{\dh }Y_{3ijk}^{1} &=&\text{\dh }\overline{\text{\dh }}%
Y_{3ijk}^{2}=-10Y_{3ijk}^{2} \\
\text{\dh }Y_{3ijk}^{0} &=&-12Y_{3ijk}^{1} \\
\text{\dh }\overline{\text{\dh }}Y_{3ijk}^{2} &=&-10Y_{3ijk}^{2} \\
\overline{\text{\dh }}\text{\dh }Y_{3ijk}^{1} &=&-10Y_{3ijk}^{1} \\
\overline{\text{\dh }}\text{\dh }Y_{3ijk}^{0} &=&-12Y_{3ijk}^{0}
\end{eqnarray}

\subsection{Clebsch-Gordon Expansions}

We present a table of products involving functions with $s=(2,1,0,-1,-2)$
and $l=(0,1,2).$

\subsubsection{Products of $l=1$ with $l=1:$}

\begin{eqnarray}
Y_{1i}^{1}Y_{1j}^{0} &=&\frac{i}{\sqrt{2}}\epsilon _{ijk}Y_{1k}^{1}+\frac{1}{
2}Y_{2ij}^{1} \\
Y_{1i}^{1}Y_{1j}^{-1} &=&\frac{1}{3}\delta _{ij}-\frac{i\sqrt{2}}{4}\epsilon
_{ijk}Y_{1k}^{0}-\frac{1}{12}Y_{2ij}^{0} \\
Y_{1i}^{0}Y_{1j}^{0} &=&\frac{2}{3}\delta _{ij}+\frac{1}{3}Y_{2ij}^{0}
\end{eqnarray}

\subsubsection{Products of $l=1$ with $l=2:$}

\begin{eqnarray}
Y_{1i}^{1}Y_{2jk}^{2} &=&Y_{3ijk}^{3} \\
Y_{1i}^{0}Y_{2jk}^{0} &=&-\frac{4}{5}\delta _{kj}Y_{1i}^{0}+\frac{6}{5}
(\delta _{ij}Y_{1k}^{0}+\delta _{ik}Y_{1j}^{0})+\frac{1}{5}Y_{3ijk}^{0} \\
Y_{1i}^{1}Y_{2jk}^{0} &=&\frac{2}{5}Y_{1i}^{1}\delta _{kj}-\frac{3}{5}
Y_{1j}^{1}\delta _{ik}-\frac{3}{5}Y_{1k}^{1}\delta _{ij}  \nonumber \\
&&+\frac{i}{\sqrt{2}}(\epsilon _{ikl}Y_{2jl}^{1}+\epsilon _{ijl}Y_{2kl}^{1})+%
\frac{2}{5}Y_{3ijk}^{1} \\
Y_{1i}^{1}Y_{2jk}^{1} &=&-\frac{1}{6}\text{\dh }[Y_{1i}^{1}Y_{2jk}^{0}] \\
Y_{2ij}^{-1}Y_{1k}^{1} &=&\frac{3}{10}Y_{1i}^{0}\delta _{jk}+\frac{3}{10}
Y_{1j}^{0}\delta _{ik}-\frac{2}{10}Y_{1k}^{0}\delta _{ij}  \nonumber \\
&&+\frac{i\sqrt{2}}{12}[\epsilon _{jkl}Y_{2il}^{0}+\epsilon _{ikl}Y_{2lj}^{0}
]-\frac{1}{30}Y_{3ijk}^{0} \\
Y_{1i}^{0}Y_{2jk}^{1} &=&-\frac{2}{5}Y_{1i}^{1}\delta _{jk}+\frac{3}{5}
Y_{1j}^{1}\delta _{ik}+\frac{3}{5}Y_{1k}^{1}\delta _{ij}  \nonumber \\
&&-\frac{i}{3\sqrt{2}}(\epsilon _{ikl}Y_{2jl}^{1}+\epsilon
_{ijl}Y_{2kl}^{1})+\frac{4}{15}Y_{3ijk}^{1} \\
Y_{2ij}^{2}Y_{1k}^{-1} &=&\frac{3}{10}\delta _{jk}Y_{1i}^{1}+\frac{3}{10}
\delta _{ik}Y_{1j}^{1}-\frac{1}{5}\delta _{ij}Y_{1k}^{1}  \nonumber \\
&&-\frac{i\sqrt{2}}{12}[\epsilon _{ikl}Y_{2jl}^{1}+\epsilon
_{jkl}Y_{2il}^{1} ]-\frac{1}{30}Y_{3ijk}^{1} \\
Y_{2ij}^{2}Y_{1k}^{0} &=&\text{\dh }[Y_{2ij}^{2}Y_{1k}^{-1}]
\end{eqnarray}

\subsubsection{Products of $l=2$ with $l=2:$}

\textbf{total} $s=4$\newline

\begin{equation}
Y_{4ijkl}^{4}=Y_{2ij}^{2}Y_{2kl}^{2}
\end{equation}

\textbf{total} $s=3$\newline
\begin{equation}
Y_{2kl}^{2}Y_{2ij}^{1}=-\frac{i}{\sqrt{2}}[\epsilon
_{ile}Y_{3jke}^{3}+\epsilon _{jke}Y_{3ile}^{3}]+\frac{1}{2}Y_{4ijkl}^{3}
\end{equation}

\textbf{total} $s=2$\newline
The numbers in parentheses as superscripts give the eigenvalues of the
associated quantities:

\begin{eqnarray}
Y_{2kl}^{1}Y_{2ij}^{1} &=&\frac{3}{7}K_{ijkl}^{2(0)}+\frac{4}{7}
K_{ijkl}^{2(14)}, \\
Y_{2kl}^{0}Y_{2ij}^{2} &=&-\frac{3}{7}K_{ijkl}^{2(0)}+\frac{1}{2}{K}
_{ijkl}^{2(6)}+\frac{3}{7}K_{ijkl}^{2(14)},
\end{eqnarray}
where the $K^{\prime }s$ are eigenfunctions with $l=(2,3,4)$%
\begin{eqnarray}
K_{ijkl}^{2(0)} &=&Y_{2kl}^{1}Y_{2ij}^{1}-\frac{2}{3}%
(Y_{2kl}^{0}Y_{2ij}^{2}+Y_{2kl}^{2}Y_{2ij}^{0}),\qquad l=2 \\
K_{ijkl}^{2(6)} &=&Y_{2kl}^{0}Y_{2ij}^{2}-Y_{2kl}^{2}Y_{2ij}^{0},\qquad
\qquad \qquad \qquad \;l=3 \\
K_{ijkl}^{2(14)} &=&Y_{2kl}^{1}Y_{2ij}^{1}+\frac{1}{2}%
(Y_{2kl}^{0}Y_{2ij}^{2}+Y_{2kl}^{2}Y_{2ij}^{0}),\qquad l=4
\end{eqnarray}
and explicitly decomposed as 
\begin{eqnarray}
K_{ijkl}^{2(0)} &=&-\frac{8}{3}(\delta _{kl}Y_{2ij}^{2}+\delta
_{ij}Y_{2kl}^{2})+2[\delta _{lj}Y_{2ik}^{2}+\delta _{ik}Y_{2lj}^{2}+\delta
_{kj}Y_{2il}^{2}+\delta _{il}Y_{2kj}^{2}],  \nonumber \\
K_{ijkl}^{2(6)} &=&\frac{i}{\sqrt{2}}[\epsilon _{ike}Y_{3ejl}^{2}+\epsilon
_{jke}Y_{3eil}^{2}+\epsilon _{ile}Y_{3ejk}^{2}+\epsilon _{jle}Y_{3eik}^{2}],
\nonumber \\
K_{ijkl}^{2(14)} &=&\frac{1}{2}Y_{4ijkl}^{2},
\end{eqnarray}

\textbf{total} $s=1$\newline

\begin{eqnarray}
Y_{2ij}^{0}Y_{2kl}^{1} &=&\frac{3}{10}J_{4ijkl}^{1(0)}+\frac{1}{14}
J_{4ijkl}^{1(4)}+\frac{1}{5}J_{4ijkl}^{1(10)}+\frac{3}{7}J_{4ijkl}^{1(18)} \\
Y_{2ij}^{-1}Y_{2kl}^{2} &=&\frac{1}{20}J_{4ijkl}^{1(0)}+\frac{1}{28}
J_{4ijkl}^{1(4)}-\frac{1}{20}J_{4ijkl}^{1(10)}-\frac{1}{28}J_{4ijkl}^{1(18)}
\end{eqnarray}
where the J's are eigenfunctions with $l=(1,2,3,4)$%
\begin{eqnarray}
J_{4ijkl}^{1(0)}
&=&C_{4ijkl}^{1}-D_{4ijkl}^{1}+4E_{4ijkl}^{1}-4F_{4ijkl}^{1}\,,\,\,\,\,\,\,%
\,\;\;l=1 \\
J_{4ijkl}^{1(4)}
&=&C_{4ijkl}^{1}+D_{4ijkl}^{1}+12E_{4ijkl}^{1}+12F_{4ijkl}^{1}\,\,,\,\;l=2 \\
J_{4ijkl}^{1(10)}
&=&C_{4ijkl}^{1}-D_{4ijkl}^{1}-6E_{4ijkl}^{1}+6F_{4ijkl}^{1}\,,\,\,\,\,\,\,%
\;\;l=3 \\
J_{4ijkl}^{1(18)}
&=&C_{4ijkl}^{1}+D_{4ijkl}^{1}-2E_{4ijkl}^{1}-2F_{4ijkl}^{1}\,\,,\,\,\,\,\,%
\,\;\;l=4
\end{eqnarray}
with 
\begin{eqnarray}
C_{4ijkl}^{1} &\equiv &Y_{2ij}^{0}Y_{2kl}^{1} \\
D_{4ijkl}^{1} &\equiv &Y_{2ij}^{1}Y_{2kl}^{0} \\
E_{4ijkl}^{1} &\equiv &Y_{2ij}^{-1}Y_{2kl}^{2} \\
F_{4ijkl}^{1} &\equiv &Y_{2ij}^{2}Y_{2kl}^{-1}
\end{eqnarray}
and decomposed as

\begin{eqnarray}
J_{4ijkl}^{1(0)} &=&i2\sqrt{2}[\delta _{ik}\epsilon _{ljf}+\delta
_{jk}\epsilon _{lif}+\delta _{il}\epsilon _{kjf}+\delta _{jl}\epsilon _{kif}
]Y_{1f}^{1},  \nonumber \\
J_{4ijkl}^{1(4)} &=&6[\delta _{li}Y_{2kj}^{1}+\delta _{ik}Y_{2jl}^{1}+\delta
_{jk}Y_{2il}^{1}+\delta _{jl}Y_{2ik}^{1}]-8[\delta _{ij}Y_{2lk}^{1}+\delta
_{kl}Y_{2ij}^{1}],  \nonumber \\
J_{4ijkl}^{1(10)} &=&-\frac{i}{\sqrt{2}}[\epsilon
_{ike}Y_{3ejl}^{1}+\epsilon _{jke}Y_{3eil}^{1}+\epsilon
_{ile}Y_{3ejk}^{1}+\epsilon _{jle}Y_{3eik}^{1}],  \nonumber \\
J_{4ijkl}^{1(18)} &=&\frac{1}{3}Y_{4ijkl}^{1},
\end{eqnarray}

\textbf{total }$s=0$\newline

\begin{eqnarray}
Y_{2kl}^{2}Y_{2ij}^{-2} &=&\frac{1}{5}F_{ijkl}^{0(0)}+\frac{2}{5}
F_{ijkl}^{0(2)}+\frac{2}{7}F_{ijkl}^{0(6)}+\frac{1}{10}F_{ijkl}^{0(12)}+%
\frac{1}{70}F_{ijkl}^{0(20)}, \\
Y_{2kl}^{1}Y_{2ij}^{-1} &=&\frac{4}{5}F_{ijkl}^{0(0)}+\frac{4}{5}
F_{ijkl}^{0(2)}-\frac{4}{7}F_{ijkl}^{0(6)}-\frac{4}{5}F_{ijkl}^{0(12)}-\frac{%
8}{35}F_{ijkl}^{0(20)}, \\
Y_{2kl}^{0}Y_{2ij}^{0} &=&\frac{24}{5}F_{ijkl}^{0(0)}-\frac{48}{7}
F_{ijkl}^{0(6)}+\frac{72}{35}F_{ijkl}^{0(20)},
\end{eqnarray}
with the eigenfunctions $F$ with $l=(0,1,2,3,4)$

\begin{eqnarray}
l &=&0:F_{ijkl}^{0(0)}=(Y_{2kl}^{2}Y_{2ij}^{-2}+Y_{2ij}^{2}Y_{2kl}^{-2})+%
\frac{1}{4}(Y_{2kl}^{1}Y_{2ij}^{-1}+Y_{2ij}^{1}Y_{2kl}^{-1})+\frac{1}{24}
Y_{2kl}^{0}Y_{2ij}^{0}, \\
l &=&1:F_{ijkl}^{0(2)}=(Y_{2kl}^{2}Y_{2ij}^{-2}-Y_{2ij}^{2}Y_{2kl}^{-2})+%
\frac{1}{8}(Y_{2kl}^{1}Y_{2ij}^{-1}-Y_{2ij}^{1}Y_{2kl}^{-1}), \\
l &=&2:F_{ijkl}^{0(6)}=(Y_{2kl}^{2}Y_{2ij}^{-2}+Y_{2ij}^{2}Y_{2kl}^{-2})-%
\frac{1}{8}(Y_{2kl}^{1}Y_{2ij}^{-1}+Y_{2ij}^{1}Y_{2kl}^{-1})-\frac{1}{24}
Y_{2kl}^{0}Y_{2ij}^{0}, \\
l &=&3:F_{ijkl}^{0(12)}=(Y_{2kl}^{2}Y_{2ij}^{-2}-Y_{2ij}^{2}Y_{2kl}^{-2})-%
\frac{1}{2}(Y_{2kl}^{1}Y_{2ij}^{-1}-Y_{2ij}^{1}Y_{2kl}^{-1}),\qquad \text{
\qquad }\qquad \\
l
&=&4:F_{ijkl}^{0(20)}=(Y_{2kl}^{2}Y_{2ij}^{-2}+Y_{2ij}^{2}Y_{2kl}^{-2})-(Y_{2kl}^{1}Y_{2ij}^{-1}+Y_{2ij}^{1}Y_{2kl}^{-1})+%
\frac{1}{4}Y_{2kl}^{0}Y_{2ij}^{0},
\end{eqnarray}
and decomposed into

\begin{eqnarray}
F_{ijkl}^{0(0)} &=&\frac{1}{2}\delta _{ik}\delta _{jl}+\frac{1}{2}\delta
_{li}\delta _{kj}-\frac{1}{3}\delta _{ij}\delta _{kl},  \nonumber \\
F_{ijkl}^{0(2)} &=&i\frac{\sqrt{2}}{8}(\delta _{jl}\epsilon _{ike}+\delta
_{ki}\epsilon _{jle}+\epsilon _{jke}\delta _{il}+\delta _{kj}\epsilon
_{ile})Y_{1e}^{0},  \nonumber \\
F_{ijkl}^{0(6)} &=&\frac{1}{6}(\delta _{ij}Y_{2kl}^{0}+\delta
_{kl}Y_{2ij}^{0})-\frac{1}{8}(\delta _{lj}Y_{2ik}^{0}+\delta
_{ki}Y_{2lj}^{0}+\delta _{li}Y_{2kj}^{0}+\delta _{kj}Y_{2il}^{0}),  \nonumber
\\
F_{ijkl}^{0(12)} &=&-\frac{i}{24\sqrt{2}}[\epsilon
_{ike}Y_{3ejl}^{0}+\epsilon _{jke}Y_{3eil}^{0}+\epsilon
_{ile}Y_{3ejk}^{0}+\epsilon _{jle}Y_{3eik}^{0}],  \nonumber \\
F_{ijkl}^{0(20)} &=&\frac{1}{24}Y_{4ijkl}^{0}.
\end{eqnarray}

\end{document}